\documentstyle[aps]{revtex}

\begin{document}
\draft
\title{Rigidly rotating cylinders of charged dust}
\author{B.V.Ivanov\thanks{%
E-mail: boyko@inrne.bas.bg}}
\address{Institute for Nuclear Research and Nuclear Energy,\\
Tzarigradsko Shausse 72, Sofia 1784, Bulgaria}
\maketitle

\begin{abstract}
The gravitational field of a rigidly rotating cylinder of charged dust is
found analytically. The general and all regular solutions are divided into
three classes. The acceleration and the vorticity of the dust are given as
well as the conditions for the appearance of closed timelike curves.
\end{abstract}


\section{Introduction}

Uncharged dust with spherical symmetry exists only as a flat spacetime. The
addition of charge allows the existence of non-trivial solutions
representing the electrification of Minkowski spacetime \cite{one,two,three}%
. On the contrary, when the symmetry is cylindrical, static charged dust
configuration are still trivial. One needs at least rigid rotation to invoke
charged analogs of the Lanczos-Van Stockum metric \cite{four,five}. A
solution with vanishing Lorentz force was found in Ref. \cite{six}. A
thorough study of this problem was performed in the works of Islam \cite
{seven,eight,nine,ten,eleven,twelve,thirteen,fourteen}. The emphasis,
however, was on the axially symmetric rigid rotation. The charge-mass ratio
was taken to be constant. Solutions with non-constant ratio were also found 
\cite{fifteen}, one of them being the general solution with vanishing
Lorentz force. Later the general solution of this type, when the charged
dust is axisymmetric and differentially rotating, was derived \cite
{sixteen,seventeen,eighteen}. Several cylindrically symmetric stationary
solutions were obtained as particular cases.

In this paper we study from the beginning cylindrically symmetric rigidly
rotating configurations of charged dust. The general and all regular
solutions are found, they fall into three distinct cases. Connections are
made with past results.

In Sec II the Einstein-Maxwell equations are written for the Papapetrou form
of the metric. A master differential equation is given for $g_{tt}$ and all
other dust characteristics are written in terms of this metric component.
Expressions are given for the acceleration, vorticity and angular velocity.
Criteria for elementary flatness and the appearance of closed timelike
curves (CTC) are discussed. In Sec III a symmetry transformation is used to
break the main equation into three different cases and their general
solutions are given. They resemble the vacuum solution with its Weyl and
Lewis classes. In Sec IV all regular solutions are derived and studied in
more detail. They are appropriate for interior solutions. Connections with
previous research are outlined. Sec V contains a short discussion.

\section{Field equations and dust characteristics}

The metric of a stationary cylindrically symmetric spacetime has the
Papapetrou form \cite{nineteen,twenty} 
\begin{equation}
ds^2=-f\left( dt+Ad\varphi \right) ^2+f^{-1}\left[ e^{2k}\left(
dr^2+dz^2\right) +W^2d\varphi ^2\right] ,  \label{one}
\end{equation}
where $x^0=t,x^1=\varphi ,x^2=r,x^3=z$ and the metric functions depend only
on the radial coordinate. Dust has zero pressure so that $W=r$ and the
Einstein equations read 
\begin{equation}
R_{\mu \nu }-\frac 12g_{\mu \nu }R=8\pi T_{\mu \nu },  \label{two}
\end{equation}
\begin{equation}
T_{\mu \nu }=\rho u_\mu u_\nu +E_{\mu \nu }.  \label{three}
\end{equation}
We use units with $G=c=1$. In comoving coordinates the four-velocity of dust 
$u^\mu $ has just a time component $u^0=f^{-1/2}$. The energy density is $%
\rho =mn$ where $m$ is the mass of the particles and $n$ is their number
density. Furthermore 
\begin{equation}
8\pi E_{\mu \nu }=2F_\mu ^\alpha F_{\nu \alpha }-\frac 12g_{\mu \nu
}L,\qquad L=F_{\alpha \beta }F^{\alpha \beta },  \label{four}
\end{equation}
\begin{equation}
F_{\mu \nu }=A_{\mu ,\nu }-A_{\nu ,\mu },  \label{five}
\end{equation}
where $A_\mu $ is the electromagnetic vector potential and derivatives are
denoted by comma. In our case 
\begin{equation}
A_\mu =\left( \Phi ,\Psi ,0,0\right)   \label{six}
\end{equation}
and there is radial electric field and longitudinal magnetic field. The
covariant Maxwell equations are 
\begin{equation}
\left( F^{\mu \nu }\right) _{;\nu }=4\pi J^\mu ,\qquad J^\mu =qnu^\mu
=\alpha \rho u^\mu ,  \label{seven}
\end{equation}
where $q$ is the charge of the particles and $\alpha =q/m$. There are two
non-trivial equations 
\begin{equation}
\Psi ^{\prime }-A\Phi ^{\prime }=\frac{br}f,  \label{eight}
\end{equation}
\begin{equation}
4\pi \alpha \rho =-\frac{f^{3/2}e^{-2k}}r\left( \frac{r\Phi ^{\prime }}f%
+bA\right) ^{\prime }.  \label{nine}
\end{equation}
Here $b$ is an integration constant, specific for the charged case, and $%
^{\prime }$ means derivative with respect to $r$. The electromagnetic
invariant $L$ reads 
\begin{equation}
L=\frac 12e^{-2k}\left( b^2-\Phi ^{\prime 2}\right) .  \label{ten}
\end{equation}
We use the same combinations of Ricci tensor components as in the neutral
case \cite{nineteen}. The Einstein-Maxwell equations become 
\begin{equation}
\frac{f^2A^{\prime }}r=a_0+4b\Phi ,  \label{eleven}
\end{equation}
\begin{equation}
f\left( f^{\prime \prime }+\frac{f^{\prime }}r\right) -f^{\prime 2}=-\frac{%
f^4A^{\prime 2}}{r^2}+8\pi f\rho e^{2k}+2f\left( b^2+\Phi ^{\prime 2}\right)
,  \label{twelve}
\end{equation}
\begin{equation}
\frac{2k^{\prime }}r-\frac{f^{\prime 2}}{2f^2}+\frac{f^2A^{\prime 2}}{2r^2}=%
\frac 2f\left( b^2-\Phi ^{\prime 2}\right) ,  \label{thirteen}
\end{equation}
\begin{equation}
2k^{\prime \prime }+\frac{f^{\prime 2}}{2f^2}+\frac{f^2A^{\prime 2}}{2r^2}=%
\frac 2f\left( b^2+\Phi ^{\prime 2}\right) ,  \label{fourteen}
\end{equation}
where $a_0$ is a constant. The Lorentz force is given by 
\begin{equation}
m\left( \frac{du^\mu }{ds}+\Gamma _{\nu \alpha }^\mu u^\nu u^\alpha \right)
=qF^{\mu \nu }u_\nu   \label{fifteen}
\end{equation}
and vanishes when $\Phi ^{\prime }=0$. Its $r$ component supplies a relation
between $f$ and $\Phi $%
\begin{equation}
f=\left( b_0-\alpha \Phi \right) ^2,  \label{sixteen}
\end{equation}
where $b_0$ is an arbitrary constant. A system of 6 independent equations
for the 6 variables $f,A,k,\rho ,\Phi ,\Psi $ is given by Eqs
(8,9,11-13,16). The charge $q$ and the mass $m$ are free parameters.

One can write the r.h.s. of Eq (11) as 
\begin{equation}
a_0+4b\Phi =b_1-\frac{4b}\alpha \sqrt{f},\qquad b_1=a_0+\frac{4bb_0}\alpha ,
\label{seventeen}
\end{equation}
where $b_1$ is a redefinition of $b_0$. Inserting Eqs (9,11,16,17) into Eq
(12) we obtain an ordinary differential equation for $f$%
\begin{equation}
\left( q^2-m^2\right) \left( f^{\prime \prime }+\frac{f^{\prime }}r-\frac{%
f^{\prime 2}}f\right) +2\left( 4m^2-q^2\right) b^2-\frac{6mqbb_1}{\sqrt{f}}+%
\frac{q^2b_1^2}f=0.  \label{eighteen}
\end{equation}

The other characteristics of the charged dust are expressed through $f$.
Thus, $\Phi $ is given by Eq (17), $A$ is obtained by integrating Eq (11) 
\begin{equation}
A=b_1\int \frac{rdr}{f^2}-\frac{4b}\alpha \int \frac{rdr}{f^{3/2}}.
\label{nineteen}
\end{equation}
Then $\Psi $ and $\rho $ are obtained from Eqs (8,9) which become 
\begin{equation}
\Psi ^{\prime }=\frac{br}f-\frac{Af^{\prime }}{2\alpha f^{1/2}},
\label{twenty}
\end{equation}
\begin{equation}
4\pi \alpha \rho e^{2k}=\frac{f^{\prime \prime }}{2\alpha }+\frac{f^{\prime }%
}{2\alpha r}-\frac{3f^{\prime 2}}{4\alpha f}-\frac{bb_1}{f^{1/2}}+\frac{4b^2}%
\alpha .  \label{twentyone}
\end{equation}
Finally, $k$ follows from Eq (13) 
\begin{equation}
\frac{2k^{\prime }}r=\frac{\alpha ^2-1}{2\alpha ^2}\frac{f^{\prime 2}}{f^2}-%
\frac{b_1^2}{2f^2}+\frac{4bb_1}{\alpha f^{3/2}}+\frac{2\left( \alpha
^2-4\right) b^2}{\alpha ^2f}.  \label{twentytwo}
\end{equation}

The characteristics of the four-velocity in the metric (1) are of special
interest \cite{twentyone}. The expansion and shear vanish. The acceleration
has only a radial component $v^r$ and its magnitude is 
\begin{equation}
v=\left( \sqrt{f}\right) ^{\prime }e^{-k}=-\alpha \Phi ^{\prime }e^{-k}.
\label{twentythree}
\end{equation}
The vorticity vector has only a longitudinal component $w^z$ and magnitude 
\begin{equation}
w=\frac{A^{\prime }}{2r}f^{3/2}e^{-k}=\left( \frac{b_1}{2f^{1/2}}-\frac{2b}%
\alpha \right) e^{-k}.  \label{twentyfour}
\end{equation}

The condition for elementary flatness is \cite{twenty,twentyone} 
\begin{equation}
\lim\limits_{r\rightarrow 0}\frac{e^{-k}}r\left( r^2-f^2A^2\right) ^{1/2}=1.
\label{twentyfive}
\end{equation}
We shall show that the second term in the bracket decouples for regular
solutions so that $k\left( 0\right) =1$ follows. Then the angular velocity $%
\Omega =w\left( 0\right) $ of the rigid body rotation becomes 
\begin{equation}
\Omega =\frac{b_1}{2f\left( 0\right) ^{1/2}}-\frac{2b}\alpha .
\label{twentysix}
\end{equation}

The appearance of CTC is governed by $g_{11}$%
\begin{equation}
g_{11}=f^{-1}\left( r+fA\right) \left( r-fA\right) .  \label{twentyseven}
\end{equation}
The sign of $A$ determines the direction of rotation. For definiteness we
accept that $A>0$. The condition for CTC is $g_{11}<0$ which means 
\begin{equation}
A>\frac rf.  \label{twentyeight}
\end{equation}

The key equation (18) was obtained by Islam \cite{fourteen}, Eq (6.102) in
the Lewis formulation of the problem. We have rederived it in the Papapetrou
picture because it is simpler and draws parallels with uncharged rotating
perfect fluids with $\gamma $-law equation of state \cite{twentyone}. Eq
(18) is quite complicated and seems non-integrable. Different particular
cases were studied by Islam. In the next section we simplify considerably
this equation and find its general solution.

\section{General solution}

The derivation of Eq (18) makes it possible to trace the appearance of the
constants $b,b_1$. The potential $\Phi $ of the electric field is defined up
to a constant, $\Phi \rightarrow \Phi +\Phi _0$. Under this transformation
Eqs (16,17) give 
\begin{equation}
b_0\rightarrow b_0+\alpha \Phi _0,\qquad b_1\rightarrow b_1+4b\Phi _0.
\label{twentynine}
\end{equation}
In Ref. \cite{fourteen} $\Phi _0$ was used to nullify $b_0$. However, we can
nullify $b_1$ instead, as long as $b\neq 0$. Hence, either $b=0$ or $b_1=0$.
Eq (18) simplifies drastically. This mechanism does not work when $\Phi $ is
constant and only magnetic field is present. Then the Lorentz force vanishes
and $f$ is also a constant. In conclusion, there are three cases

1) $\Phi =0,$ $f=f_0$. Eq (18) turns into a quadratic equation for $%
f_0^{1/2} $%
\begin{equation}
2\left( 4m^2-q^2\right) b^2f_0-6mqbb_1\sqrt{f_0}+q^2b_1^2=0.  \label{thirty}
\end{equation}
The case $q=\pm m$ eliminates the derivatives of $f$ in Eq (18) and again
leads to Eq (30).

2) $b=0$. Eq (18) becomes 
\begin{equation}
\left( \frac{rf^{\prime }}f\right) ^{\prime }=\lambda _2\frac r{f^2},\qquad
\lambda _2=\frac{q^2b_1^2}{2\left( m^2-q^2\right) }.  \label{thirtyone}
\end{equation}

3) $b_1=0$. Eq (18) now reads 
\begin{equation}
\left( \frac{rf^{\prime }}f\right) ^{\prime }=2\lambda _3\frac rf,\qquad
\lambda _3=\frac{\left( 4m^2-q^2\right) b^2}{m^2-q^2}.  \label{thirtytwo}
\end{equation}

The last two equations are written together in a simple manner if we
introduce $f=e^{2u}$%
\begin{equation}
\left( ru^{\prime }\right) ^{\prime }=\lambda _ire^{-n_iu}.
\label{thirtythree}
\end{equation}
Here $i=2,3$ and $n_2=4$, $n_3=2$. One can easily see that in the neutral
vacuum case ($\rho =\Phi =\Psi =b=0$) Eqs (11,12) give exactly Eq (31) or
(33) with $i=2$ and negative $\lambda _2$. This is in fact Eq (A4) from Ref. 
\cite{twentytwo} which leads to the three classes of Lewis solutions \cite
{twentythree,twentyfour,twentyfive}. Thus the charged dust solutions are
similar to the uncharged vacuum solutions, but as we shall see, for
realistic situations $q^2<m^2$ and $\lambda _2$ is positive.

Eq (33) is solved by introducing $s=\ln r$ and $y=n_iu-2\ln r$. It becomes 
\begin{equation}
y_{ss}=n_i\lambda _ie^{-y}  \label{thirtyfour}
\end{equation}
and is integrated after multiplication with $y_s$. The general solution is 
\begin{equation}
f_i=\left( ar^{\sqrt{c}+1}+\frac{n_i\lambda _i}{8ac}r^{-\sqrt{c}+1}\right)
^{4/n_i},  \label{thirtyfive}
\end{equation}
where $a$ and $c$ are integration constants. The function $f$ depends on
four parameters: $a,c,\alpha $ and either $b$ or $b_1$. The constant $c$ is
real and when $c>0$ Eq (35) is analogous to the vacuum Weyl class. When $c$
is negative, $c=-\beta ^2$, $f$ can be kept real if $a$ becomes complex and
satisfies 
\begin{equation}
\left| a\right| ^2=-\frac{n_i\lambda _i}{8\beta ^2}.  \label{thirtysix}
\end{equation}
A necessary condition is $\lambda _i<0$. This requires $\alpha ^2>1$ for $%
i=2 $ and $1<\alpha ^2<4$ for $i=3$. Then we can write $a=\left| a\right|
e^{i\sigma }$ and 
\begin{equation}
f_i=\left[ 2\left| a\right| r\cos \left( \sigma +\beta \ln r\right) \right]
^{4/n_i}.  \label{thirtyseven}
\end{equation}
This is an analog of the vacuum Lewis class. The free parameters are $\sigma
,\beta ,\alpha $ and $b_1\left( b\right) $.

The other characteristics follow from $f$ and are real too. The integrations
in Eq (22) are not explicit in general. Curiously, in the case $b=0$ the
function $A$ may be written as in the vacuum 
\begin{equation}
A=\frac{b_1r^{\sqrt{c}+1}}{2a\sqrt{c}}+A_0,  \label{thirtyeight}
\end{equation}
where $A_0$ becomes a complex constant in the Lewis class.

The vacuum solutions are exterior ones and the singularities on the axis do
not matter. The rotating charged dust is supposed to serve as an interior
solution, so the study of regular metrics is really important.

\section{Regular solutions}

{\it Case 1. Solutions with vanishing Lorentz force.} They have $\Phi =0$
and constant $f$ given by Eq (30) 
\begin{equation}
\sqrt{f_0}=\frac{b_1\alpha \left( 3\pm \sqrt{1+2\alpha ^2}\right) }{2b\left(
4-\alpha ^2\right) },  \label{thirtynine}
\end{equation}
unless $q^2=4m^2$. The special case $q^2=m^2$ gives the simple result 
\begin{equation}
f_0=\frac{2\pm \sqrt{3}}6\left( \frac{b_1}b\right) ^2.  \label{forty}
\end{equation}
In the case $q^2=4m^2$ Eq (30) is linear and 
\begin{equation}
f_0=\frac 19\left( \frac{b_1}b\right) ^2  \label{fortyone}
\end{equation}
if $b_1\neq 0$. When $b_1$ vanishes, $f_0$ remains unfixed. The other
functions read 
\begin{equation}
A=\frac{a_0r^2}{2f_0^2},\qquad \Psi =\frac{br^2}{2f_0},\qquad a_0=b_1-\frac{%
4b}\alpha \sqrt{f_0},  \label{fortytwo}
\end{equation}
\begin{equation}
k=\frac{k_0r^2}4,\qquad k_0=\frac{2b^2}{f_0}-\frac{a_0^2}{2f_0^2},
\label{fortythree}
\end{equation}
\begin{equation}
\rho =-\frac{ba_0}{4\pi \alpha \sqrt{f_0}}e^{-2k},\qquad L=\frac{b^2}2%
e^{-2k}.  \label{fortyfour}
\end{equation}
Obviously $A$ decouples from the elementary flatness condition Eq (25). The
solution has three parameters; $\alpha ,b,b_1$. The constant $b$ sets the
strength of the magnetic field and $ba_0/\alpha $ must be negative to ensure
the positivity of the energy density. The latter decreases monotonically
when $k_0$ is positive, but never vanishes.

One can recover the Lanczos solution by setting $b=-a_0q/2\rightarrow 0$ and 
$f_0=1$, since Eq (30) becomes trivial. Then $k_0$ is always negative and $%
\rho $ increases with the distance. The presence of magnetic field invokes $%
b $ in $k_0$ and makes it positive when the rotationally induced $a_0$ is
compensated.

Eq (23) shows that there is no acceleration of the dust particles. The
angular velocity and the vorticity are 
\begin{equation}
\Omega =\frac{a_0}{2\sqrt{f_0}},\qquad w=\Omega e^{-k}.  \label{fortyfive}
\end{equation}
CTC appear when $r$ is bigger than 
\begin{equation}
r_0=\frac{\sqrt{f_0}}\Omega .  \label{fortysix}
\end{equation}
The solution is regular and realistic. It was found in Ref. \cite{six}, see
also Eq (6.96) from Ref. \cite{fourteen}, where $f_0=1$ was chosen.

Turning to the other two cases, Eq (37) shows that $f$ from the Lewis class
is always singular at the axis, while Eq (35) indicates that the Weyl class
is regular only for $c=1$. For simplicity, let us take $a=n_i\lambda _i/8$,
so that $f\left( 0\right) =1$. Eq (35) becomes 
\begin{equation}
f_i=\left( 1+\frac{n_i\lambda _i}8r^2\right) ^{4/n_i},  \label{fortyseven}
\end{equation}
which is the general regular solution for non-constant $f$.

{\it Case 2: }$b=0$. This case is close to the vacuum solution and we have 
\begin{equation}
f_2=1+\frac{\lambda _2}2r^2,  \label{fortyeight}
\end{equation}
\begin{equation}
A=\frac{b_1r^2}{2f_2},\qquad \Psi ^{\prime }=\frac{b_1r}{2\alpha f_2^{3/2}},
\label{fortynine}
\end{equation}
\begin{equation}
4\pi \rho =\frac{\lambda _2\left( 4-\lambda _2r^2\right) }{4\alpha ^2f_2}%
e^{-2k},\qquad e^{-2k}=f_2^{\frac{1-\alpha ^2}{\alpha ^2}},  \label{fifty}
\end{equation}
\begin{equation}
\Phi ^{\prime }=-\frac{\lambda _2r}{2\alpha f_2^{1/2}},\qquad L=-\frac 12%
\Phi ^{\prime 2}e^{-2k},  \label{fiftyone}
\end{equation}
\begin{equation}
v=\frac{\lambda _2}{b_1}rw,\qquad w=\Omega f_2^{\frac{1-2\alpha ^2}{2\alpha
^2}},\qquad \Omega =\frac{b_1}2.  \label{fiftytwo}
\end{equation}
The function $A$ again decouples from Eq (25). Eq (50) shows that $\rho
\left( 0\right) >0$ when $\lambda _2>0$ ($q^2<m^2$). Then, however, $e^{-2k}$
monotonically increases and so does $L$. The acceleration also increases,
while the vorticity increases when $q^2<m^2/2$ and decreases when $%
m^2/2<q^2<m^2$. The energy density has a zero at $r_0=2/\sqrt{\lambda _2}$
and turns negative, so the junction to an exterior should be made before
this point. According to Eq (28) CTC exist when $\Omega r>1$. They will
appear in the interior if $q^2<\frac 23m^2$. Some of the above relations
were found in the Lewis form of the metric by Islam \cite{fourteen}.

{\it Case 3; }$b_1=0$. Then $\lambda _3$ is positive except for $%
m^2<q^2<4m^2 $ and we have 
\begin{equation}
f_3=\left( 1+\frac{\lambda _3}4r^2\right) ^2,  \label{fiftythree}
\end{equation}
\begin{equation}
A=\frac{4b\left( 1-f_3\right) }{\alpha \lambda _3f_3},\qquad \Psi ^{\prime
}=\left( b-\frac{\lambda _3}{2\alpha }\right) \frac r{f_3}+\frac{\lambda _3r%
}{2\alpha },  \label{fiftyfour}
\end{equation}
\begin{equation}
4\pi \rho =\left[ \frac{b^2\left( 8m^2-5q^2\right) }{\alpha ^2\left(
m^2-q^2\right) }-\frac{\lambda _3^2}{4\alpha ^2}r^2\right] e^{-2k},\qquad
e^{-2k}=f_3^{\frac{2\left( 1-\alpha ^2\right) }{\alpha ^2}},
\label{fiftyfive}
\end{equation}
\begin{equation}
\Phi ^{\prime }=-\frac{\lambda _3}{2\alpha }r,\qquad L=\frac 12\left( b^2-%
\frac{\lambda _3^2}{4\alpha ^2}r^2\right) e^{-2k},  \label{fiftysix}
\end{equation}
\begin{equation}
v=-\frac{\alpha \lambda _3}{4b}rw,\qquad w=\Omega f_3^{\frac{1-\alpha ^2}{%
\alpha ^2}},\qquad \Omega =-\frac{2b}\alpha .  \label{fiftyseven}
\end{equation}
We accept $\Omega >0$ and $q>0$, so that $b<0$. We have $A>0$ and again $%
A\sim r^2$ for small $r$, decoupling in Eq (25). Now there are two intervals
where $\rho \left( 0\right) $ is positive, $q^2<m^2$ and $q^2>\frac 85m^2$.
The function $e^{-2k}$ increases with the distance when $q^2<4m^2$ and
decreases when $q^2>4m^2$. In the second case $\rho $ also decreases
monotonically. This shows that realistic solutions are possible even when $%
q^2>m^2$ contrary to the assertions in Ref. \cite{fourteen}. In any case the
density has a zero and changes sign at 
\begin{equation}
r_0^2=\frac{4\left( \alpha ^2-1\right) \left( 5\alpha ^2-8\right) }{\left(
\alpha ^2-4\right) ^2b^2}.  \label{fiftyeight}
\end{equation}

Finally, CTC exist when the inequality 
\begin{equation}
\Omega r\left( 1+\frac{\lambda _3}8r^2\right) >1  \label{fiftynine}
\end{equation}
is satisfied. It is slightly more complicated than in the previous two
cases. CTC appear in the interior solution (i.e. before $r_0$) if 
\begin{equation}
4\left( 3+\gamma \right) \left( 12+5\gamma \right) \left( 12+7\gamma \right)
^2>\gamma ^4\left( 4+\gamma \right) .  \label{sixty}
\end{equation}
Here $\alpha ^2=4+\gamma $. This is true for small $\gamma $ but for big
positive $\gamma $ the r.h.s. dominates. The situation is more intricate for
negative $\gamma $ where the inequality does not hold around the zeroes of
the polynomial in the l.h.s.

{\it Case 3a: }$b_1=0,$ $q^2=4m^2.$ Then $\lambda _3=0$ and the previous
solution belongs also to Case 1 with $f_0=1$, $k_0=0$. The energy density
and $L$ are constant. However, the general formula in Eq (35) gives also
another solution, singular at the origin 
\begin{equation}
f=c_1r^{c_2},  \label{sixtyone}
\end{equation}
where $c_i$ are some constants. It leads to 
\begin{equation}
A=c_3r^{2-3c_2/2},\qquad k=\frac 3{16}c_2^2\ln r,  \label{sixtytwo}
\end{equation}
\begin{equation}
\Phi ^{\prime }=-\frac{\sqrt{c_1}c_2}4r^{c_2/2-1},\qquad \Psi ^{\prime }=%
\frac b{c_1}\left( 1+\frac{c_2}2\right) r^{1-c_2}.  \label{sixtythree}
\end{equation}
This solution was studied in Refs. \cite{eight,eleven,twelve} where $\rho $
can be found too. The importance of a general solution, encompassing both
cases was addressed there. We would like to mention also that $\Psi \sim
r^{2-c_2}$ except for $c_2=2$ when $\Psi \sim \ln r$. This particular case
was studied in Ref. \cite{twelve} where erroneously $\Psi $ was taken as a
constant.

\section{Discussion}

The master equation (18) of the problem studied in this paper has been known
for a long time. It looks formidable, non-integrable and even more
complicated than the Painlev\'e equations and transcendents, that appear in
the electrovac case \cite{twentysix}. However, one of the two main constants
in it can be always set to zero by changing the electric potential. This
allows to obtain the general solution, which breaks into three distinct
cases. One is the peculiar case with vanishing Lorentz force. The neutral
dust solution of Lanczos can be obtained from it by taking a specific limit.
The other two cases resemble the vacuum solution and posses both Weyl and
Lewis classes of solutions. CTC appear already in the Weyl class.

The general solutions are singular at the origin. We have derived in
addition all regular metrics which serve as interior charged dust solutions.
They have been found in the past by Islam but as particular cases of a
supposedly non-integrable equation. Here they have been systematized in an
exhausting classification scheme, some corrections were made and their
properties were further studied. We have stressed the advantages of the
Papapetrou form of the metric for this and related problems \cite
{nineteen,twentyone,twentytwo}. As usual, the Weyl-Papapetrou-Majumdar type
of connection given by Eq (16) makes the problem analytically solvable in
elementary functions, unlike the electrovacuum case. Matching these interior
charged dust solutions to third Painlev\'e transcendents seems too
complicated, although some concrete cases are known \cite{fourteen}.

\end{document}